\documentclass[nonacm,anonymous=false]{acmart}

\usepackage[disable]{todonotes}
\usepackage{hyperref}
\usepackage{cleveref}
\usepackage{pdfpages}
\usepackage{natbib}

\makeatletter
\renewcommand{\todo}[2][]{%
  \@todo[fancyline,caption={#2}, #1]{
    \begin{minipage}{\linewidth}
      \footnotesize #2
    \end{minipage}
  }%
}
\makeatother
\newcommand{\michiin}[1]{\todo[color=yellow!40,inline]{\textbf{Michi}: #1}}

\newcommand{\eg}{e.g.\@ }

\begin{document}

%% Don't Touch ! %%%%%% 
%\coverpage           % 
%\linenumbers         % 
%\doublespacing       % 
%%%%%%%%%%%%%%%%%%%%%%% 

%%% ALL TEXT OF ARTICLE BELOW THIS LINE %%%
%-----------------------------------------%
%%%%%%%%%%%%%%%%% TITLE %%%%%%%%%%%%%%%%%%%%%%%%%%%%
\title{Learning Classifier Systems for Self-Explaining Socio-Technical-Systems}
\titlenote{This article is an extension of previous work \citep{Heider2021} presented as part of \emph{LIFELIKE 2021 co-located with the 2021 Conference on Artificial Life ({ALIFE} 2021)}}

%%%%%%%%%%%%%%%% AUTHORS %%%%%%%%%%%%%%%%%%%%%%%%%%%
% For each author, associate with affiliations below.
%   Comma-separate authors.
%   Twitter handle is optional 
%   (for journal's promotional purposes).
\author{Michael Heider}
\email{Michael.Heider@uni-a.de}
\orcid{0000-0003-3140-1993}
\affiliation{%
  \institution{Universit\"at Augsburg}
  \streetaddress{Am Technologiezentrum 8}
  \city{Augsburg}
  \country{Germany}
}
\author{Helena Stegherr}
\email{Helena.Stegherr@uni-a.de}
\orcid{0000-0001-7871-7309}
\affiliation{%
  \institution{Universit\"at Augsburg}
  \streetaddress{Am Technologiezentrum 8}
  \city{Augsburg}
  \country{Germany}
}
\author{Richard Nordsieck}
\email{Richard.Nordsieck@xitaso.com}
\affiliation{%
  \institution{Xitaso GmbH}
  \streetaddress{Austraße 35}
  \city{Augsburg}
  \country{Germany}
}
\author{J\"org H\"ahner}
\email{Joerg.Haehner@uni-a.de}
\orcid{0000-0003-0107-264X}
\affiliation{%
  \institution{Universit\"at Augsburg}
  \streetaddress{Am Technologiezentrum 8}
  \city{Augsburg}
  \country{Germany}
}

%%%%%%%%%%%%%%% ABSTRACT %%%%%%%%%%%%%%%%%%%%%%%%%%%
% Type your abstract below
\begin{abstract}
In socio-technical settings, operators are increasingly assisted by decision support systems. 
By employing these, important properties of socio-technical systems such as self-adaptation and self-optimization are expected to improve further.
To be accepted by and engage efficiently with operators, decision support systems need to be able to provide explanations regarding the reasoning behind specific decisions.
In this paper, we propose the usage of Learning Classifier Systems, a family of rule-based machine learning methods, to facilitate transparent decision making and highlight some techniques to improve that.
We then present a template of seven questions to assess application-specific explainability needs and demonstrate their usage in an interview-based case study for a manufacturing scenario.
We find that the answers received did yield useful insights for a well-designed LCS model and requirements to have stakeholders actively engage with an intelligent agent.
\end{abstract}

%%%%%%%%%%%%%% KEYWORDS %%%%%%%%%%%%%%%%%%%%%%%%%%%%%%
% Please provide 5 to 6 keywords, comma-separated

\maketitle
\section{Introduction}

\begin{figure}
	\centering
	\includegraphics[width=.62\linewidth]{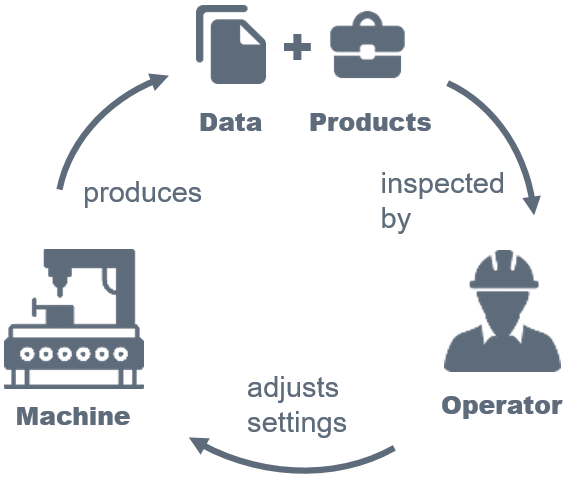}
	\caption{Operator-in-the-loop in today's productions.}%
	\label{fig:intro-today}
\end{figure}

Increasing automation of manufacturing creates a continuous interest in properties commonly associated with lifelike or organic computing systems, such as self-adaptation or self-optimisation, within the producing industry~\citep{permin2016}.
These properties are often achieved using data driven and learning methods~\citep{zhang2017,lughofer2019,schoettler2019}, as with increasing digitalisation and IoT efforts, data can be collected in large amounts.
In modern factories, products are usually inspected by the machines' operators or specialized quality assurance personnel to assess their quality, cf.\@~\Cref{fig:intro-today}.
For the sake of simplicity, we subsume both roles under the term `operator'.
Recent advances in automated inspection often integrate computer vision-based approaches  \citep{markgraf2017}.
However, these can be of limited use when quality is not assessable from the surface, e.g.\@ structural or chemical properties that involve laboratory testing.
Thus, these systems currently can only partially automate inspection while the conclusions with regards to machine reconfiguration are still reached manually. 
This requires a large amount of operator knowledge and experience to achieve optimal or even satisfactory results.
In settings with heterogeneous machines and few operators, the strain on operator experience is further increased and production can be seriously threatened by a loss of qualified personnel, \eg through retirement.

\begin{figure}
	\centering
	\includegraphics[width=.75\linewidth]{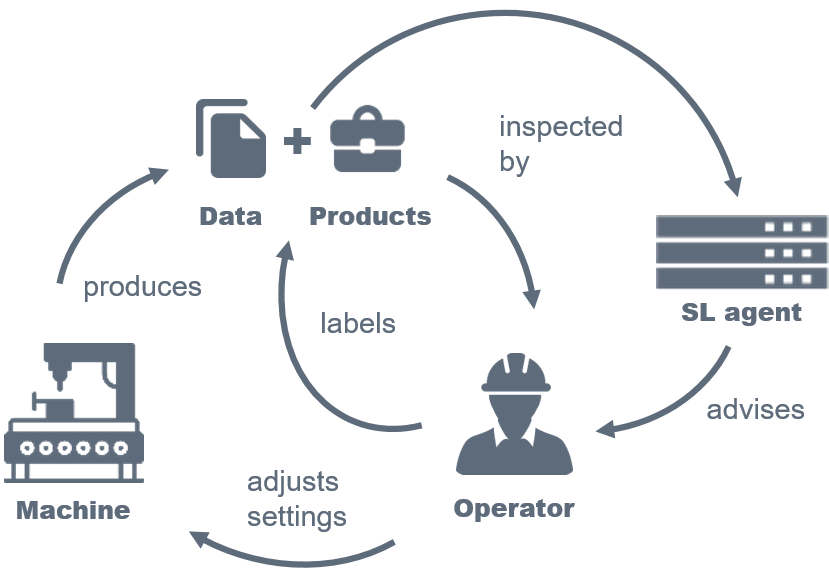}
	\caption{Assisted production using an agent trained with supervised learning (SL) during operation.}%
	\label{fig:intro-sl}
\end{figure}

To reduce reliance on specific knowledge of operators and improve the self-adapting and self-optimizing systems, the operator can be assisted by decision support systems. % can as an intermediate step 
These can easily incorporate large amounts of information simultaneously and are less biased to well known settings, especially compared to operators that only have limited understanding of or experience with the machines.
Such decision support systems utilize learning from past experience and ongoing human expert feedback.
Combining human operators and supervised learning (SL) agents that collaboratively adjust machines (or lines thereof) that manufacture products expands the socio-technical system with a collaborative decision making dimension, cf.\@~\Cref{fig:intro-sl}.

Typical shopfloor environments will feature many workers operating on many machines, but not necessarily in a one to one array, \eg multiple workers might be needed to operate a single machine while multiple other machines can be operated by a single worker due to automation.
Additionally, to utilize the available data most efficiently, not every machine should need its own model, but models should generalise over multiple machines of the same or similar type.
For production lines where multiple models would participate, the parametrization choices of preceding machines would need to be accounted for by subsequent models, \eg through the help of models of higher abstraction.
In this environment, each individual model takes input from and advise multiple operators, while each individual operator might interact with different models throughout a shift.

An integral element for implementing these systems is that operators are able to trust decisions made by their recommendation agents.
This requires the system to be self-explaining in both adequate form and abstraction level.
It involves an explanation regarding the basis of the recommendation, \eg what input parameters led to this output, as well as an assessment of the quality of the decision, \eg what is the expected error in quality when executing the recommended parametrization.
In this paper, we posit that Learning Classifier Systems are well-suited to be used within the proposed SL agent by reviewing different explainability techniques in light of this setting. 
We then introduce a template of research questions that need to be addressed to successfully apply LCSs (or other rule-based systems) in this context.
We demonstrate the successful usage of those questions in a case study where we utilize them in a sequence of interviews with stakeholders from a producing company, the REHAU SE.

\section{Learning Classifier Systems}

\emph{Learning Classifier Systems} (LCSs) are a family of rule-based learning systems \citep{urbanowicz2009}.
While LCSs are a diverse field, they share some common properties. 
In general, LCSs produce models consisting of a finite number of if-then rules (traditionally often referred to as \emph{classifiers}), where individual premises (\emph{conditions}) and, by extension, the global model structure, are optimized using a---typically evolutionary---metaheuristic and the \emph{conclusions} of the rules use a problem-dependent model.
These classifiers or local models can then individually be ascribed a quality of their prediction within their respective subspace of the global model's input space.
In our view, this structure is sufficient to motivate their application within a decision support system.
However, we acknowledge that as there is a plethora of possibilities to train such a model, choosing the `right' LCS for an actual implementation needs to be done use-case specific, as some LCSs will yield better performing models than others and their transparency varies.

\subsection*{Explainability in LCSs}

Explainability of machine learning models is usually differentiated into
\begin{itemize}
\item \emph{transparent methods}, allowing interpretation of decisions and comprehension of the model based on the model structure itself, and 
\item \emph{post-hoc methods}, utilising visualisation, transformation of models into intrinsically transparent models and similar techniques on models that are not by themselves transparent \citep{barredoArrieta2020}.
\end{itemize}
As rule-based learning systems, LCSs generally fall into the domain of transparent models and are regarded as excellent for interpretability due to their relation to human behaviour.
However, several factors can limit the degree to which humans can easily comprehend the model and follow its decision making process. 
Most notable are the number of classifiers and their formulation.
Conditions of rules in complex feature spaces are harder to understand than those that operate directly on the data, \eg higher level features aggregating multiple sensor readings versus the readings themselves.
Additionally, conditions can be formulated using non-linear functions rather than readable decision boundaries \citep{bull2002}.
Conclusions of rules that utilize complex black box models, such as neural networks \citep{lanzi2006}, are also harder to understand than linear or constant models, even if these local black box models are usually much smaller than a model of the same class that encompasses the complete problem space would need to be.

These issues can warrant design adjustments within the LCS or the application of post-hoc methods.
The number of rules can be combated depending on the type of system considered: For Pittsburgh-style systems, this is usually achieved by promoting small individuals through adjustments of the fitness function \citep{Bacardit2007}, whereas in Michigan-style systems, rule subsumption and compaction methods are applied \citep{tan2013,liu2019,liu2021b}.
An improved understanding of singular classifiers can be pursued by promoting simplicity during training through a suitable fitness function, by applying analyses typical for the respective models, \eg feature importance estimations in neural networks, and with a variety of visualisation methods \citep{urbanowicz2012,liu2019,liu2021}.

\section{LCSs in Industrial Decision Support Systems}

Many different LCSs have been proposed over the years and while originally envisioned as a powerful reinforcement learner, they have been extended for all learning paradigms \citep{urbanowicz2009}.
We consider the application as a decision support system that proposes settings to an operator and informs them of the reasoning behind this choice to be an SL task.
This can be solved with either online or offline learning as long as the model used to make recommendations provides a compacted version of itself for inference and subsequently serving explanations.
The LCS learns from experiences including sensor readings, product information, used machine settings and resulting quality measures, all of which will be a mixture of real and categorical values.
When tasked with assisting an operator, the SL agent uses sensor readings and product information to propose machine settings and predict the expected quality.

Besides the previously introduced explainability techniques, LCSs also easily allow us to provide operators with all examples from our training data that formed the local model (as we know which examples were matched by the rule's condition).
This can help further the trust that the model's predictions are actually based on existing expertise.
Going beyond traditional explaining by example \citep{barredoArrieta2020}, each example that influenced an individual rule's weights could theoretically be listed, whereas in black-box models usually the entire sample influences every weight.

In Michigan-style LCSs, each individual rule gets ascribed a quality measure (or multiple thereof in XCS(F)).
This (or in case of multiple measures, at least one of them) represents the rule's fitness and is used to guide an evolutionary process.
Moreover, we can utilize these measures to provide our operator with additional information on how exact and therefore useful a recommendation is.
Rules with a low prediction quality and thus a high expected error might provide poor machine settings, while other rules in the model might actually provide very useful settings.
This disparity in different parts of the feature space can also allow insights into where new sampling should take place \citep{stein2017} and allows to differentiate the model further.
Even if---viewed globally---the model is less than optimal, it can still be used within the SL agent and aid operators on problem instances where it is well fit.

\section{A Template to Assess Explainability Requirements and LCS Model Design}
\label{questions}

Following this theoretical examination of the applicability of LCSs as decision support systems for the parametrization of industrial machinery in a complex socio-technical environment, we want to raise several research questions that---in our view---need to be answered on a case by case basis.
We assume that some parallels will exist between applications, it seems, however, unlikely that general answers will hold for all or even the majority of cases.
Note that we broaden the scope from our operators that interact directly or indirectly with the machine to all stakeholders that have a vested interest in the operation of the shopfloor, both digital and analogue.
Thus, this could also include regulatory bodies, safety officers, engineering, management, customers and others.

\paragraph{To what extent does a stakeholder request explanations?}  % if changed, also change in concl.
This can have numerous dimensions, such as depth, frequency or diversity of explanations.
Someone that operates the machine directly might prefer examples of past experiences while quality assurance personnel might prefer visualisations or vice versa. 
In this question we assume that stakeholders may seek explanations that go beyond regulatory requirements, although a potential answer may be that they are not interested in further/deeper explanations.
This raises another aspect: How important is explainability deemed if prediction quality potentially suffers?

\paragraph{What are the differences within a type of stakeholder?} 
Tying directly into the previous question, we assume that the diverse stakeholders of a given type will answer questions regarding explainability differently. 
Individual stakeholders may also hold different understandings of the machine itself, so explanations would need to accommodate specific levels of prior knowledge.
Furthermore, diversity between individual operators might be substantial and warrant personalisation approaches.

\paragraph{How many rules may the served model contain before being too large?} 
Smaller rule sets are easier to generate a general understanding on, while larger rule sets can provide a more diverse coverage of the input space and therefore more accurate predictions.
In some cases, like explanations for specific decisions, the entirety of the rule set might not even be of interest and stakeholders may prefer explanations to be limited to the rules whose conditions matched the situation.

\paragraph{What form can conditions take before they are too complex to be understood?} 
Many rule representations have been proposed in the past and while ellipsoids or neural networks can provide improved results, cuboids (a simple interval for each input dimension) might be easier to comprehend.
This should also probe whether the exact condition is even considered relevant or if operators are content with knowing that it applies to a certain instance.

\paragraph{How important are explanations of why the decision boundary of a rule is placed a certain way?} 
In LCSs, the model structure (and decision boundary of each rule) is optimized using a metaheuristic to localise the rules in a way that they fit the data well. 
Within this question, we want to ascertain how important insights into this process are to operators.

\paragraph{What form can conclusions take before they are too complex to be understood?} 
While linear models are widely regarded as easily comprehensible, more complex models might yield better results and typical explanations, such as feature importance analysis, may satisfy the stakeholders' want for understanding the decision making process.
This also translates to the usage of mixing models (where multiple rules are used to construct a prediction) and the comprehension thereof.

\paragraph{What information do stakeholders request about the training process?}
Relating to question 5, this question aims towards the training in general and what steps are performed in the process towards deriving a model rather than at an analysis of the utilized model.
An important aspect of this can be the gathering, cleaning and selection of data and responsibilities therein.

\section{Case Study: Assisting Operators in a Chemical Industry Plant}

To demonstrate a potential use of our proposed template of questions to determine the requirements for a self-explaining socio-technical system that supports operators in their day to day tasks while also satisfying other stakeholders needs, we performed an interview-based case study.
In this case study we interviewed a variety of different stakeholders about their individual as well as their colleagues' and subordinates' needs for such a system before its final design and implementation.
The envisioned operator assistance system (OAS) is to be employed in an international chemical industry company, the REHAU SE \footnote{https://www.rehau.com/}.
REHAU plans on piloting it in a German plant of their interior solutions branch which is the main focus of our case study, where so called \emph{edge bands} are produced. 
However, we also interviewed a stakeholder from a plant of their window solutions branch to broaden the scope, potentially find differences even between branches of a single company and, hopefully, find some answers that can be applied to other branches in the future as well.

\subsection{Operator Assistance System}

The primary motivation behind the operator assistance system (OAS) is to disencumber operators and reduce their overall workload, which currently is substantial.
This is to be done through increased automation of, currently manual, routine adjustments and by providing operators with more insights into disturbances and with potential solutions.
Overall, this increases the robustness of the production and reduces material and energy waste.

In the line control OAS like systems assist operators at manual configuration and partially automates it.
Its components are largely well understood from a chemical engineering point of view.
While arguably some level of explanations to operators could always be beneficial, these algorithms do not employ any form of machine learning component and therefore fall out of scope for our study, where the focus is an SL agent operating as one of the, potentially many, systems forming the overall OAS. 
Another component currently in production is a tool that aggregates existing knowledge in an easy to navigate tree-like structure. 
When encountering some issue, e.g.\@ a quality defect, the operator navigates via web interface from broad areas to specific defects/disturbances where individual stages are described both textually and visually. 
Once the issue is narrowed down, the operator is presented with common solutions to the problem and an estimate on how successful these have been in the past. 
After the issue is resolved, the operator is asked to give feedback whether the provided suggestion was helpful and correct, promoting this suggestion for the next operator that encounters this issue.
These paths through the tree-like structure can be reformulated as rules.
Those rules can potentially in turn be used inside an LCS, either as an initial population before training or by manual insertion into the trained model, where they serve to cover areas of the problem space where training examples were too scarce to create sufficiently accurate rules.
Additionally, these rules can be used for potential explanations of evolved rules, as they should---due to their crowd-sourced nature---be deemed more reliable by operators than some rather abstract machine learning process. 
For simplicities sake, we refer to the envisioned SL-based agent as part of the OAS as the \emph{agent} in the remainder of this text and primarily consider its specific requirements without limiting other components. 

Depending on its maturity, predictive power and stakeholder trust, the agent can be employed at different levels:
\begin{enumerate}
	\item Predict the quality of a machine parametrization selected by the operator,
	\item actively make suggestions for possible parametrizations and their predicted product quality to the operator, 
	\item set a single parametrization and prompt the operator to confirm and
	\item regulate the process parameters fully automatically, e.g.\@ when product quality or process stability indicators drop, with the operator only acting as a supervisor.
\end{enumerate}
These levels also change the operator's role in our socio-technical system of machine, agent and operator in that the higher levels lessen the mental load of trying to come up with possible solutions and transform the operator to an executor of physical adjustments and tasks while keeping him in a position of supervisory responsibility.
Likely, different settings in which the agent is to be used will allow higher levels of operation earlier.
In less crucial or sensitive, e.g.\@ prone to significant damage, parts of a production line, the agent will be able to choose from a wider range of still sufficient parametrizations while facing less scrutiny by different stakeholders.
The same holds for areas with different data availability and quality. 
Ultimately, any SL prediction is dependent on diverse and correct data for training.
Machines of a line that have long been digitised and fitted with well calibrated sensors will more likely offer such data than machines that have until recently been controlled by analogue means.
For these newly digitised machines, it might even be unknown what sensors are missing to make meaningful predictions and they might not yet have been online long enough to gather sufficient data or even to allow the determination of what noise is to be expected during operation, e.g.\@ the impact of seasonal changes.

Regardless of the specific scenarios, it is clear that to get such an agent into production, relevant stakeholders have to be on board from the early stages of its design process.
This was also reflected by those stakeholders in early talks about potential use cases.
In these talks they first raised the, albeit expected, issue of transparency of such an agent and its decisions as central towards generating enough trust to employ it.

\subsection{Extrusion: An example application of the OAS and its agent}
\label{extrusion_example}

In the production of plastics, a typical first part of a production line is the melting of synthetic granulates (or powders) and subsequent form-giving extrusion of the heated semi-fluid mass.
The correct pressure---and for many products also the temperature---is crucial to ensure sufficient dimensional accuracy and therefore product quality.
The exact values are primarily dependent on size, shape and material type, but from a process engineering point of view it is very much possible to find a range of values that can be considered sufficiently optimal to guarantee the desired product quality.
Operators will control for this measurable parameter rather than shape and size itself, as process engineering guarantees desired dimensionality whenever the correct pressure was applied.
This also has one key advantage for prediction: 
The resulting learning task is a regression for which sensor readings are comparatively easy to obtain, whereas the control of a multidimensional shape and size vector for which complicated and highly accurate laser scans would be needed is much less straightforward.

In REHAU's interior solutions branch, specifically edge band production, extrusion pressure is regulated by eight adjustable parameters.
Additionally, a multitude of additional sensor readings, primarily temperatures in different sections of the extruder, are available.
The adjustable parameters show highly non-linear relationships with the target, warranting sophisticated self-learning and---due to the requirements on transparency---self-explaining systems.

\subsection{Study Design}

One critical issue to be solved to actually get the agent into use in a scenario similar to the one presented in \Cref{extrusion_example} is stakeholder acceptance.
This acceptance needs to be nurtured from the early design stages by making choices according to wishes (and worries) of the various stakeholders.
From early preliminary talks with R\&D and different management levels, we already knew that whatever the exact embedding system design will be, the self-explainability of the employed agent is likely central.
This already hinted towards an LCS being a very plausible choice for the learning algorithm.
Thus, we use the template raised in \Cref{questions} with relevant stakeholders to determine if the assumptions that explainability is very important are even correct and, if so, how the resulting LCS model should likely be designed. 
This serves a second purpose, as discussing these issues in form of the questions with the stakeholders allows them early participation in the design process and can be used to develop according to their requirements.
This reinforces the perception of holding a stake rather than the feeling that some ill-suited system was forced onto them.
In another direction but complementary to the described goals this also facilitates a test of the validity and applicability of the questions raised and whether they even allow meaningful insights.
This is an important consideration for potential future applications of the template (or if they turn out to be suboptimal for a reformulated version).

To validate the applicability of the questions and to gain some perspective on what answers we can expect and where additional clarifications or input might be warranted, we conducted a pilot interview with the Director of Smart Factory and selected members of his department, which is responsible for machine automation, data science, IoT, assistance systems and sensors.
We found that the questions can be used as proposed in \Cref{questions}, but some more explanations, especially into the specific nature of LCS models, are beneficial to get more useful answers for the LCS-specific questions.
Importantly, we found that explanations are definitely sought after on many levels.
More results are discussed in \Cref{pilotinterview}.

Our main study is conducted in individual interviews with stakeholders of about 45 minutes.
As all participants were German and few work with English on a daily basis, these interviews were conducted in German.
Interviews began with the interviewee prompted to give some information about themselves and their current job as well as job history at REHAU.
After a short introduction into the general topic, possible levels on which the agent can operate and an example use case based on the extruder (cf. \Cref{extrusion_example}), the stakeholders were presented with the seven questions and some additional explanations, examples and follow-ups.
The questions were also reformulated into German and technical (machine learning) jargon was---where possible---kept to a minimum.
As LCS (and other ML model types) were unknown to most participants, an example of a 1-D task solution and an eight-dimensional example rule were also presented before question 3, where the number of rules is discussed.
Interviewees were strongly encouraged to ask for clarifications if some point of a question was unclear and received some additional context or details if they expressed trouble answering.
The interview was aided by a set of slides, so interviewees could read along and reread the question if needed.
These slides can be found at \url{https://doi.org/10.5281/zenodo.6505010}.

The relevant archetypical stakeholder roles can be summarized as follows:
\begin{itemize}
\item \emph{Operators} operate the machine to manufacture a product. 
Typically, operation takes place in a one to one ratio in the interior solutions branch and sometimes in a one to many ratio in the window solutions branch. 
They interact with the agent throughout their shift and, as they are responsible for smooth production, rely heavily on its capabilities.
Especially (comparatively) inexperienced operators often need assistance, whereas seasoned (10+ years of experience) operators will rarely be in situations where they consult others.
\item \emph{Team Leaders} supervise a group of operators on the shop floor within a given shift.
For troubleshooting, team leaders are the subsequent responder when the colleague on the next line was unable to assist an operator.
Therefore, they interact with the line control (and thus the OAS and the agent) on a frequent basis.
If even some of the operators' questions and issues get resolved by an agent, the team leaders' job becomes considerably less stressful, while if the agent gives poor advise or confuses the operator, their job might become more difficult.
\item \emph{Production Managers} are ultimately responsible for the entire plants production and are thus very interested in past and projected manufacturing capabilities.
\item \emph{Process Engineers} have the deepest knowledge of the underlying process. 
They have deep foundational understanding towards maintaining process stability, which they are also constantly trying to improve. 
They operate either closely to/within the plant, where they are second in line for troubleshooting when operators and their supervisors could not fix the issue at hand by themselves, or in more centralised process engineering departments, where they determine set parameters and machine configurations for new material compositions and product types and perform other developmental steps towards machine improvement and innovation.
\item \emph{Data Scientists} are expected to maintain, improve and expand the capabilities and possible applications of the agent (and other machine learning methods in use).
They select which data is to be used, what new sensors are needed and validate the correctness of readings.
From a model perspective, it is likely that the agent encapsulates multiple models that are directly trained to predict on this machine (model) or even for a specific product, rather than a singular generalized model that solves all tasks, although generalization is overall desirable as fewer models are easier to maintain. 
Data scientists would thus need to determine which machines and products can share a model and for which combinations other models are needed.
Ultimately, a badly performing agent is the responsibility of the data science team. 
\end{itemize}

These stakeholder archetypes are often not clearly distinguished in a single person and their view on certain aspects might be heavily influenced by their (job) history so that despite their current position, they still express views we can clearly attribute to another archetype. 
%\michi{wir haben auch eine nicht klare operative Trennung dieser Rollen}
Team leaders are often trained process engineers that have been operators at REHAU before undergoing additional education.
In-plant process engineers often have a management role as well, with responsibilities for sections of the plant.
Although, in this archetype specifically, the exact position of a person between R\&D responsibilities, where university graduates are more common, and day-to-day operations widely varies.
We still chose to present these as one archetype as the general questions they ask of the agent are similar.
The interview partners available for this study were selected to allow an overview of all roles and interviewees were asked to distinguish between the different roles they might find themselves in or have held in the past for their answers.
They were also requested to separately answer for operators and based on their perception on operator's requirements.

\subsection{Interview Findings}

From the conducted interviews we find that there is a need for self-explainability of the envisioned agent and that easier models are generally preferable.
More detailed descriptions of the answers to the seven questions are shown in \Cref{pilotinterview} for our pilot interview and \Cref{maininterview} for the main interviews that were conducted afterwards.
Reassuringly, we also found that the agent is indeed wanted.

\subsubsection{Pilot Interview}
\label{pilotinterview}

In this first interview we primarily aimed at validating the applicability of the proposed questions to gain insights regarding the envisioned scenario.
It was conducted with the Director of Smart Factory at REHAU in the presence of some members of his department, who were involved in the already existing parts of the OAS and a variety of data science applications.
The concept of a SL-based agent to assist operators and its various levels of application were well established previously.
Answers, as given by the Director regarding his perception of various stakeholders' requirements, were recorded and are stated in the following:

\paragraph{To what extent does a stakeholder request explanations?}

For operators, this primarily depends on the autonomy of the agent.
The more autonomous the agent acts, the less will the individual operator request explanations. 
In contrast, the process engineer will always want in-depth explanations. 
This requirement will likely increase with agent autonomy, e.g.\@ when debugging potential issues, as the operator will have less insights into what was configured and why.
Team leaders will require more explanations and more depth than operators.
Data scientist will want maximal transparency and self-explainability.

\paragraph{What are the differences within a type of stakeholder?}

For operators the frequency and depth of explanations will highly depend on their experience. 
Experienced operators will probably disregard the agent completely and use their own knowledge to solve upcoming issues.
Thus, they will also not request any explanations.
For other stakeholders, experience might matter for simple tasks, e.g.\@ if the prediction aligns with their mental model, they will not request an explanation, but overall explanations will be requested by all personnel in these roles.

\paragraph{How many rules may the served model contain before being too large?}

As LCS models and their structure's implications were not completely clear, we presented a small ad-hoc visual aid what an LCS model might look like, which we then also kept for the main interviews.
Data scientists may be the only stakeholders that might want to analyse the model in its entirety. 
Other stakeholders, specifically operators and team leaders, will be primarily interested in the model's situation-specific predictions.
Therefore, explanations of the given mixing model will be more relevant and the global model can contain a large quantity of rules as long as it can still be experimentally or statistically verified, i.e.\@ through well-chosen test data.
The mixing model should also contain few rules.
This question also brought up a point about submodels: 
They should be trained in a way as to directly determine the most important features/parameters for a given prediction, e.g.\@ by forcing 2-3 weights to be considerably larger than others in a linear model.

\paragraph{What form can conditions take before they are too complex to be understood?}

Interval-based rule conditions are strongly preferred.

\paragraph{How important are explanations of why the decision boundary of a rule is placed a certain way?}

Operators and team leaders will probably not have this question and take the conditions as is.
Some trace-back to the training sample might be interesting for data scientists but is not needed.

\paragraph{What form can conclusions take before they are too complex to be understood?}

This question was deemed impossible to answer without taking the model's task and performance into account.
In general, simpler submodels are preferred.

\paragraph{What information do stakeholders request about the training process?}

The Director was unable to confidently provide deeper insights into this question.
Likely, information is of interest but the exact levels would need to be answered by the respective stakeholders.

\subsubsection{Main Interviews}
\label{maininterview}

Following the findings of the pilot interview, an expanded introduction into both the possible application of the agent as well as LCS was prepared.
Additionally, the seven questions were translated into German and, where applicable, follow-up questions based on answers given in the pilot interview were formulated.
After that, four interviews were conducted. 
As this group of stakeholders was quite heterogeneous with different perspectives on the questions as well as operators' views, we attribute the (paraphrased) statements to the respective interviewees (A through D).
\begin{itemize}
\item \textbf{A} is currently a process engineer and supervisor with administrative responsibilities in edge band production.
They started in the company as an operator and then became team leader before the promotion into the current position.
They supervise and interact with operators and machines throughout a normal work day.
\item \textbf{B} is from the window solutions branch and head of recycling and plant optimization.
They started as an operator before training as a process engineer and receiving various promotions up to plant management. 
Therefore, they have a good perception on all relevant in-plant roles and might already give some perspective if the answers can be re-used for a similar manufacturing process for a different product at another plant.
\item \textbf{C} is head of the data lab---a department responsible for all data management, analysis and science. 
They have a strong statistics background and have been working with various stakeholders from multiple plants for years. 
This includes directly interacting with operators at the machines over long periods of time. 
\item \textbf{D} is a member of the data lab.
Originally part of R\&D, they have subsequently joined the IT department and then---with its foundation---the data lab.
They are primarily responsible for keeping data-related systems, like the envisioned agent, running and up-to-date.
\end{itemize}
In this section, the answers of the interviewees regarding the various stakeholder ar\-che\-types are presented in a question-wise manner.
Where conflicting answers were given we present both.

\paragraph{To what extent does a stakeholder request explanations?}

\begin{itemize}
\item \emph{Operator}:
New operators are thankful for all assistance, including explanations (A).
Explanations also enable them to fix issues on their own (A, B).
In general, short textual explanations of 2-3 sentences are preferred (A, B). 
Probabilities of success of a proposed parametrization and rule quality could be useful but are not mandatory as long as the model itself is not guessing (B).
Explanations should be offered on request rather than by default on every prediction/re-parametrization (B).
They could be enriched with images of issues that may arise from the suboptimal parametrization or other information about past production (A, D).
Graphs and dashboards are not useful for operators (A, B).
Neither are mathematical formulae (A, B, C, D).
As long as the performance is on some generous level of practical equivalence, transparent models are preferred over better performing ones (A).
\item \emph{Team Leader}:
In addition to textual explanations, graphs can be useful to understand and improve the manufacturing process (A).
However, as long as production proceeds as scheduled, team leaders might not care for explanations (B).
Not-as-explainable models with better performance can still be useful (A).
\item \emph{Production Manager}:
The main interest is with keeping production up and efficient (B). 
Understanding why errors are occurring is of deep interest as to prevent them in future (B). 
In addition to textual explanations, that are likely too low level for most situations in which management is involved, high-level dashboards and graphs allow them to understand their production (B).
Model transparency is more important to them, but in the end pragmatism reigns (B).
\item \emph{Process Engineer}:
Being tasked with both ad-hoc debugging and long-term improvements, process engineers have a deep interest in understanding the manufacturing process (A).
Machine learning models that may infer connections from data that are unknown or at least unquantified by humans are of great relevance to achieve their goals (A). 
However, to be analysed these models need to be as transparent as possible (A).
Diverse tools for in-depth explanations are very important (A).
Process engineers might not analyse every decision but all that went wrong, as well as the general model (A).
\item \emph{Data Scientist}:
Ideally, the model would be a complete white-box as transparency and explanations are preferable (D).
However, a substantially worse white-box model should be replaced with a gray- or black-box system that undergoes a rigorous statistical analysis (C, D).
A well-validated model that can be inspected via graphs and dashboards could be deployed even without inherent transparency (C).
Depending on the task, transparency could also be approximated via post-hoc analyses, although this would make the usability for other stakeholders questionable, depending on the correlation between the original black-box and its transfer learned pendant created through post-hoc analysis, such as LIME (C).
Explanations should be in-depth and may include mathematical formulae (C, D).
\end{itemize}

\paragraph{What are the differences within a type of stakeholder?}

For all stakeholder archetypes, substantial experience will result in some predictions and decisions being obvious, thus, not requiring explanation (A, B, C, D).
Data scientists might still want to understand how the model inferred this from data but this would not warrant a self-explaining model (C).
Less experienced stakeholders will often require more or more in-depth explanations than those with average experience, although on the other hand, very experienced stakeholders might in turn require more depth to be convinced or to understand how the model found something they did not (A, B, C).
Whether or not explanations are requested is mostly dependent on attitude and motivation rather than experience (A, B).
The broadest spectrum is shown within the operator role (A, B). 
For inexperienced operators, consulting the system replaces disturbing their colleagues and/or supervisors to ask for their help, which will increase the agent's acceptance (B).
Personalisation of explanations might be good for individual operators but would greatly complicate the team leaders' and process engineers' user experience whenever they are called for assistance (A).

%\begin{itemize}
%\item \emph{Operator}
%\item \emph{Team Leader}
%\item \emph{Production Manager}
%\item \emph{Process Engineer}
%\item \emph{Data Scientist}
%\end{itemize}

\paragraph{How many rules may the served model contain before being too large?}

\begin{itemize}
\item \emph{Operator} and \emph{Team Leader}:
The overall models number of rules can be as complex as needed, however, in a given situation only few (up to 4 (A)) may match and be included in the mixing model (A, B, C, D).
Additionally, rules should ideally be limited in a way as to promote high weights for only 3-4 parameters at most, with other parameters having considerably smaller weights (A, D).
\item \emph{Process Engineer}:
A process engineer will often analyse the full model and therefore requires it to be small (A).
However, the exact size is problem dependent (A).
For the extrusion problem, 15-20 rules should be an upper limit (A, D).
\item \emph{Data Scientist}:
Matching rules are more important than the totality of rules (C, D).
Intense validation of a subset of rules will likely allow data scientists to trust the other rules as long as they share performance metrics (C).
Overall, rule similarity is also important in that many dissimilar rules are more acceptable than high overlaps (C).
However, upon further probing, sizes of 30 to 100 rules were deemed as highly complex models for successful analysis (C, D).
\end{itemize}

\paragraph{What form can conditions take before they are too complex to be understood?}

\begin{itemize}
\item \emph{Operator}:
For operators the specific condition does not need to be analysed as long as we can assure that this rule does apply (A, B).
However, when interacting with operators to explain certain decision making processes, complex models might make this more difficult (D).
\item \emph{Team Leader} and \emph{Process Engineer}: 
Easier to analyse is preferable (B) .
They should not be more complex than intervals (A).
\item \emph{Data Scientist}:
More complex conditions should be possible as long as they undergo post-hoc analysis, e.g.\@ LIME (C).
If the LCS has proven to produce well placed decision boundaries for similar problems, not all rules of every model would need to be analysed (C).
For practically equivalent performance, easier conditions are strongly preferred (D). 
With a higher degree of automation, analysing the condition becomes more important (D).
\end{itemize}

\paragraph{How important are explanations of why the decision boundary of a rule is placed a certain way?} 

The interviewees were in agreement that there is no need for explanations why the trained model exhibits certain decision boundaries and how the optimizer found these.
The data scientist might have an interest into the process from a scientific point of view but for machine operation and operator assistance through a trained model, this is not relevant (C, D).

\paragraph{What form can conclusions take before they are too complex to be understood?}

\begin{itemize}
\item \emph{Operator}:
Operators will likely not care about specifics as long as a textual explanation for the central aspects (e.g.\@ feature importance / influence on prediction) is given (B).
Models should be linear (A). 
An analysis of the mixing of the currently matching rules is sufficient (A).
\item \emph{Team Leader} and \emph{Process Engineer}:
Submodels should be kept as simple as feasible (B).
Ideally, submodels are linear (A).
In addition to an in-production use, process engineers will also want to analyse the model(s) to improve the process itself, e.g.\@ through changes in hardware, and for this the models need to be understandable to them (D).
\item \emph{Production Manager}: 
Individual predictions are less important than overall system performance (D).
\item \emph{Data Scientist}:
The usage of more complex submodels should be possible (C, D), although simpler models are always preferable (C).
If complex models are used, they would need to undergo rigorous individual testing and analysis (C).
However, for better performing submodels, this would be worth it (C, D).
\end{itemize}

\paragraph{What information do stakeholders request about the training process?}

\begin{itemize}
\item \emph{Operator}:
The more information is available, the higher will be the operators' trust in the predictions (A). 
They care about which lines and which products were used for information gathering and by whom features were selected and models were build (A).
As long as predictions are correct, operators will take the suggestions as is and not further request such information (B).
\item \emph{Team Leader}:
More detailed information than for operators as well as some form of involvement in the design process is requested (A).
\item \emph{Production Manager}:
Some information on a high abstraction level is sufficient, e.g.\@ where did the responsibilities lie (B).
Deployment time, lifetime performance and possible adaptations based on products and performance are relevant (D).
\item \emph{Process Engineer}:
To determine if model performance is in line with the current understanding of the process, and to subsequently improve process stability on the basis of the models production, engineers require as much information as available (A, B, C).
They should also be involved early on to avoid model biases from possible correlations without causation within the data (C).
\item \emph{Data Scientist}:
While multiple stakeholder archetypes will request all information available, more than anyone else data scientists will want to do statistical testing and analysis of the models (C).
They will analyse train-test-splits in detail (C).
With the model in production, they employ statistical measures to detect possible concept or sensor drifts (C).
\end{itemize}

\subsection{Summary}
\label{studysummary}
% link this somewhere to point readers here for a quick look at the findings?

We find that stakeholder archetypes have---at times substantially---diverging requirements towards the explainability of the model.
All stakeholders would prefer transparent models as long as performance is practically equivalent.
However, should this not be the case, it highly depends on both the archetype as well as the individual person.

Within the group of operators, some might not ever consult the agent and many might not care for its explanations as long as predictions---or derived parametrizations---are correct.
Regardless, substantial numbers of operators will both listen to the agent's suggestions and check its explanations.
These explanations can serve two purposes.
On one hand, they help operators check for plausibility of a decision based on their own mental model and therefore increase trust in the agent.
On the other hand, they may update the operator's mental model which is especially important for newer and inexperienced operators that would otherwise need to rely on a colleague's or supervisor's assistance.
Regarding LCSs, operators tend to only want to analyse currently matching rules.
These should be few in number and kept as simple as possible.
Operators want explanations primarily in a short textual form, ideally, directly generated from those rules. 
With this role especially, we found staunch differences between the two plants, where interior solutions' operators want explanations much more frequently than their window solutions' counterparts, where explanations are likely only requested in case of production issues and defects.
While these might be attributable to the interviewees, it is very plausible that differences in the manufacturing process and how machines are interfaced with within the socio-technical system are the root of diverging answers, e.g.\@ the fact that within window solutions multiple lines are operated by one operator.

Team leaders largely follow the trends set by their direct subordinates, the operators.
However, due to their increased responsibilities, they require more, deeper and more diversely represented explanations from easy to analyse models.
Again, the two plants seem to differ with regards to frequency of explanations for this role, although the trend is less substantial.

Production managers will less frequently interact with the agent and primarily require information about its performance and if that is poor, will request more information into likely reasons. 
The agent could, for example, explain its poor performance in certain areas of the feature space with poor sampling, high noise or unexpectedly complex parameter-target relationships.
Individual decisions are unlikely to be analysed by production managers.
However, depending on their background, they might be quite interested in what is running in their plant and how it works from a personal motivation.

The process engineer requires the most in-depth and diverse explanations and general model analysis capabilities.
From our interviews, we found a second aspect of usage for the models besides their application within our operator assistance setting.
Namely, to analyse the models (or the agent in its entirety) to deduce process improvements that go beyond a parametrization.
This can range from the hardware setup itself to chemical mixtures of line inputs to hydraulic valve switching.
The simplest models are strongly preferred for both aspects.
Process engineers are less diverse in their requirements, both from an individual as well as a plant-wise perspective.
Contrastingly to the perception expressed by the Director of Smart Factory, not only data scientist but also process engineers will want to analyse the full model.

Data scientists were overall relatively open to deploying gray-box or even black-box models as long as they had undergone substantial statistical verification or have been made explainable through post-hoc analyses.
However, transparency is preferable as statistical verification of a black-box model can be sufficient if one deeply understands the statistical decision making process and possible fallacies therein but is hard to convey towards stakeholders that do not have such knowledge and training.

By gaining an understanding of the various stakeholders' requirements through this study we also validated that the seven questions are useful to determine them. 
We found that differences between the two plants seemingly exist for some but not all of our identified stakeholder archetypes.
Likely, different domains and companies will also yield slightly different answers.
Thus, these questions should serve as a template on how to design specific studies.
Additionally, we found that potential users not only want to be included in the agent design process but also have important uses for the agent and its models that are not included in the originally envisioned case that can, however, also be solved without a differently or separately designed system and did not come up in any other previous discussion.

\section{Conclusion}

% Old Conclusion:
In this paper, we introduced a socio-technical system within an industrial manufacturing setting where operators and SL agents collaboratively adjust machine settings to optimize product quality.
In these systems, operators can interact with a variety of heterogeneous machines and different agents throughout a single shift, while the agents also interact with different operators.
Assisting the operators with recommendations from the agents decreases the necessity for experience and helps extract and conserve experience of senior operators that might otherwise be lost over time.
We reintroduced Learning Classifier Systems (LCS) and reviewed why these rule-based systems are generally considered explainable.
Building on that, we expanded on possible requirements for the design of an LCS within our agent and highlighted beneficial properties of LCSs for this application. 
This led to the formulation of seven guiding questions to assess the explainability requirements of individual use cases.
Three of these questions are applicable to a variety of machine learning models, \eg \emph{To what extent does a stakeholder request explanations?}, and aim at analysing general wants and needs, while the other four questions are more specific for rule-based systems (LCSs, decision trees, etc.).
We expect that answers to these questions are domain- and stakeholder-specific and would need to be answered independently for each setting or even use case of similar agents.
To demonstrate the validity of the questions for gaining relevant insights, we conducted a case study at the REHAU SE company. 
We interviewed stakeholders from two manufacturing plants and the centralised data science department that offers solutions to all plants within the company.
We found that an operator assistance system utilizing machine learning would improve production.
Predictive and decision-making systems should exhibit self-explainability, although the variety, frequency and depth depend on stakeholder roles as well as individual investment. 
For some stakeholders at the second plant---even within the same company---differences were noticeable.
Thus, we got indication that a use case specific reiteration of the questions will be required.
Overall, the seven guiding questions do yield useful insights and it is possible to design a system based on these answers.
Whether such a system does actually completely fulfil all requirements---as they might not have been voiced---and if it will be well received will be answered in the future. %\richard{ich find den Satz nicht verkehrt. kann man noch gut mit nutzer-zentriertem arbeiten, agil etc kombinieren um während des Entwicklungsprozess auf initial nicht geäußerte Req agieren zu können. Wenn du richtig hardcore bist sogar mit Bildchen. Die Argumentation würde etwas die Entwicklungsprozess Ebene ergänzen. Schadet evtl nicht?}
Although we assume that general trends should be transferable, these questions and answers serve as a template whenever applying rule-based learning systems to a new scenario where comprehensibility is essential and we invite other researchers to utilize this template.
Consequently, we are confident that LCSs can introduce self-explainability into these socio-technical-systems while advancing industrial manufacturing practices.

\michiin{we should add acknowledgements here. Thanking Rehau for their cooperation and the StMWi for their partial funding}

\bibliographystyle{ACM-Reference-Format}
\bibliography{XLCS_ext}

%\appendix
%\section{User Study Slides}
%\label{slides}
%\michiin{Folien sollen nicht im appendix sondern nur im supplemental, sowie über github, veröffentlicht werden. Zur einfacheren Prüfung der Inhalte sind sie hier angefügt. Wir wollen die Folien mitveröffentlichen, damit es für Leser der Studie möglich ist auf möglicherweise durch die Formulierung der Fragen induzierte Biases zu reagieren und außerdem um die Antworten durch die gegebenen Einführungen in einen sinnvollen Kontext bezüglich fachlicher Tiefe zu setzen.}
%
%\includepdf[pages=-]{user study/slides.pdf}

\end{document}